\documentclass[useAMS,usenatbib]{mn2e}

\usepackage[T1]{fontenc}
 \usepackage{pslatex}
\usepackage{amsmath}
\usepackage{amsfonts}
\usepackage{color}
\usepackage{url}
\usepackage{graphicx}
\usepackage{subfigure}
\usepackage{amssymb}
\usepackage{soul}
\usepackage{booktabs}
\usepackage{array}
\usepackage[utf8]{inputenc}
\inputencoding{latin1}
\inputencoding{utf8}

{\newif\ifnotend
\notendtrue
\def\veclist{ABCDEFGHIJKLMNOPQRSTUVWXYZabcdefghijklmnopqrstuvwxyz.}
\def\top#1#2.{#1}
\def\tail#1#2.{#2.}
\loop\expandafter\xdef\csname v\expandafter\top\veclist\endcsname%
{{\noexpand\bf\expandafter\top\veclist}}
\edef\veclist{\expandafter\tail\veclist}
\if\veclist.\notendfalse\fi\ifnotend\repeat}
\def\e{{\rm e}}
\def\i{{\rm i}}
\def\pa{\partial}
\def\E{{\cal E}}

\mathchardef\mhyphen="2D

\title[Self-modulation of FRBs]{Self-modulation of Fast Radio Bursts}

\author[Sobacchi et al.]{Emanuele Sobacchi$^{1}$\thanks{E-mail: es3808@columbia.edu}, Yuri Lyubarsky$^2$, Andrei M. Beloborodov$^{3,4}$, Lorenzo Sironi$^1$\\
$^1$ Department of Astronomy and Columbia Astrophysics Laboratory, Columbia University, 550 West 120th Street New York, NY 10027, USA\\
$^2$ Physics Department, Ben-Gurion University, P.O.B. 653, Beer-Sheva 84105, Israel \\
$^3$ Physics Department and Columbia Astrophysics Laboratory, Columbia University, 538 West 120th Street New York, NY 10027, USA \\
$^4$ Max Planck Institute for Astrophysics, Karl-Schwarzschild-Str. 1, D-85741, Garching, Germany
}

\begin{document}

\date{}

\def\p{\partial}
\def\E{\textbf{E}}
\def\B{\textbf{B}}
\def\v{\textbf{v}}
\def\j{\textbf{j}}
\def\s{\textbf{s}}
\def\e{\textbf{e}}

\newcommand{\di}{\mathrm{d}}
\newcommand{\bfx}{\mathbf{x}}
\newcommand{\bfe}{\mathbf{e}}
\newcommand{\vlos}{\mathrm{v}_{\rm los}}
\newcommand{\Tspin}{T_{\rm s}}
\newcommand{\Tb}{T_{\rm b}}
\newcommand{\degree}{\ensuremath{^\circ}}
\newcommand{\Th}{T_{\rm h}}
\newcommand{\Tc}{T_{\rm c}}
\newcommand{\bfr}{\mathbf{r}}
\newcommand{\bfv}{\mathbf{v}}
\newcommand{\bfu}{\mathbf{u}}
\newcommand{\pc}{\,{\rm pc}}
\newcommand{\kpc}{\,{\rm kpc}}
\newcommand{\Myr}{\,{\rm Myr}}
\newcommand{\Gyr}{\,{\rm Gyr}}
\newcommand{\kms}{\,{\rm km\, s^{-1}}}
\newcommand{\de}[2]{\frac{\partial #1}{\partial {#2}}}
\newcommand{\cs}{c_{\rm s}}
\newcommand{\rb}{r_{\rm b}}
\newcommand{\rqu}{r_{\rm q}}
\newcommand{\bfOmega}{\pmb{\Omega}}
\newcommand{\bfOmegap}{\pmb{\Omega}_{\rm p}}
\newcommand{\bfXi}{\boldsymbol{\Xi}}

\maketitle

\begin{abstract}
Fast Radio Bursts (FRBs) are extreme astrophysical phenomena entering the realm of non-linear optics, a field developed in laser physics. A classical non-linear effect is self-modulation. We examine the propagation of FRBs through the circumburst environment using the idealised setup of a monochromatic linearly-polarised GHz wave propagating through a uniform plasma slab of density $N$ at distance $R$ from the source. We find that self-modulation occurs if the slab is located within a critical radius $R_{\rm crit}\sim 10^{17}(N/10^2{\rm\; cm}^{-3})(L/10^{42}{\rm\; erg\; s}^{-1}){\rm\; cm}$, where $L$ is the isotropic equivalent of the FRB luminosity. Self-modulation breaks the burst into pancakes transverse to the radial direction. When $R\lesssim R_{\rm crit}$, the transverse size of the pancakes is smaller than the Fresnel scale. The pancakes are strongly diffracted as the burst exits the slab, and interference between the pancakes produces a frequency modulation of the observed intensity with a sub-GHz bandwidth. When $R\sim R_{\rm crit}$, the transverse size of the pancakes becomes comparable with the Fresnel scale, and the effect of diffraction is weaker. The observed intensity is modulated on a timescale of ten microseconds, which corresponds to the radial width of the pancakes. Our results suggest that self-modulation may cause the temporal and frequency structure observed in FRBs.
\end{abstract}

\begin{keywords}
fast radio bursts -- radio continuum: transients -- plasmas -- instabilities -- relativistic processes
\end{keywords}


\section{Introduction}

Fast Radio Bursts (FRBs) are bright extragalactic radio flashes of millisecond duration \citep[e.g.][]{Lorimer2007, Thornton2013, Spitler+2014, Spitler2016, Petroff2016, Shannon+2018, Chime2019, Chime2019b, Chime2019a}. The high brightness temperature of FRBs suggests that they are powered by a coherent emission mechanism.

In FRBs, the electromagnetic field of the radio wave may accelerate electrons up to a significant fraction of the speed of light \citep[e.g.][]{LuanGoldreich2014}. An initially static electron will reach a speed $a_0 c$,\footnote{This is only true when $a_0\ll 1$. More generally, one can show that the maximum electron Lorentz factor is $1+a_0^2/2$ \citep[e.g.][]{GunnOstriker1971}.} where $a_0=eE_0/2\pi\nu_0 m_ec$ is the standard strength parameter of the electromagnetic wave ($E_0$ is the electric field and $\nu_0$ is the frequency of the wave). For a typical FRB, one finds that $a_0\sim 8\times 10^{-6} (\nu_0/{\rm GHz})^{-1}(L/10^{42}{\rm erg\; s}^{-1})^{1/2}(R/{\rm pc})^{-1}$, where $L$ is the isotropic equivalent of the burst luminosity and $R$ is the distance from the source. Using the fact that $L\sim 4\pi D^2S_{\nu_0}\nu_0$, where $S_{\nu_0}$ is the observed flux density and $D$ is the distance of the observer, one may finally present the strength parameter as
\begin{equation}
\label{eq:a0}
a_0 \sim 8 \times 10^{-6} \left(\frac{S_{\nu_0}}{{\rm Jy}}\right)^{1/2} \left(\frac{\nu_0}{{\rm GHz}}\right)^{-1/2} \left(\frac{D}{{\rm Gpc}}\right) \left(\frac{R}{{\rm pc}}\right)^{-1} \;.
\end{equation}
Note that $a_0\ll 1$ at the large ($R\gg 8\times 10^{-6}{\rm\; pc}$) radii that we are considering throughout the paper.

A wave propagating through an ambient medium can experience strong non-linear effects even when $a_0\ll 1$. Despite their importance for laser-plasma interaction \citep[for a review, see e.g.][]{Mourou+2006}, non-linear effects have received a limited attention from the astrophysical community \citep[in the context of FRBs, see however][]{Lyubarsky2008, Lyubarsky2018, Lyubarsky2019, Gruzinov2019, Beloborodov2020, LuPhinney2020, Lyutikov2020b, Margalit+2020, YangZhang2020}.

In this paper we focus on the self-modulation of a finite-amplitude electromagnetic wave with a wave number ${\bf k}_0$. Self-modulation occurs due to the exponential growth of two satellite waves with wave numbers ${\bf k}_0\pm{\bf k}$. The wave number $k\ll k_0$ of the electromagnetic wave intensity modulation is due to the beating of the satellite waves. The instability is excited by the non-linear component of the current at the frequency of the satellite waves. \citet[][]{Drake+1974} considered the ponderomotive force, which expels the electrons from the regions with a high intensity of radiation, as the origin of the non-linear component of the current. \citet[][]{Max+1974} included the non-linear relativistic corrections to the electron motion, but neglected the effects of the ion motion and of the thermal pressure. Both these studies considered only the case when ${\bf k}$ is either aligned or perpendicular to ${\bf k}_0$, which is a significant limitation since self-modulation naturally develops in three dimensions.

First of all, in Section \ref{sec:instabilities} we review the properties of self-modulation. We closely follow the approach of \citet[][]{Max+1974}, and extend their calculations to (i) include ion motion and thermal pressure, and (ii) examine instabilities developing for arbitrary directions of the perturbation wave vector. Readers not interested in the technical details can skip to Tables \ref{table1} and \ref{table2}, which summarise the wave vectors and the growth rates of the most unstable modes.

Then, in Section \ref{sec:FRBs} we discuss the impact of self-modulation on FRBs. We focus on the propagation of the burst through an electron-ion plasma with a non-relativistic temperature, located within a parsec from the source, as inferred from the strong and variable Faraday rotation in the repeating FRB 121102 \citep[][]{Michilli+2018}. We show that sub-bursts with a finite duration and bandwidth may be generated by self-modulation. Hence, the most prominent features of the time-frequency structure of the bursts from the repeating FRB 121102, reported by Hessels et al. (2019), may be a by-product of the FRB propagation.

Guided by the results of \citet[][]{Michilli+2018} (see in particular their Figure 6), we adopt a fiducial number density $N=10^2{\rm\; cm}^{-3}$ and a fiducial magnetic field $B = 1{\rm\; mG}$ for the electron-ion plasma.\footnote{The repeating FRB 121102 has a compact ($\lesssim 0.7{\rm\; pc}$) persistent radio counterpart \citep{Chatterjee2017, Marcote2017}, which suggests the additional presence of a relativistically hot, luminous nebula with a lower density $N\sim 1{\rm\; cm}^{-3}$ in a higher magnetic field $B\sim 60{\rm\; mG}$ \citep[][]{Beloborodov2017}. Both the hot and the cool plasma components may be present around FRB 121102. Since we consider only the effect of the FRB propagation through the cool plasma component, the effects described below do not require the presence of a hot radio nebula around the FRB source.} The electron Larmor frequency, $\omega_{{\rm L}e}=eB/m_ec\sim 2\times 10^4{\rm\; Hz}$, is smaller than the electron plasma frequency, $\omega_{{\rm P}e}=\sqrt{4\pi Ne^2/m_e}\sim 5\times 10^5{\rm\; Hz}$. Since dominant component of the non-linear current that excites the instability oscillates at twice the frequency $\omega_0$ of the electromagnetic wave \citep[e.g.][]{Max+1974}, and $\omega_0\gg\omega_{{\rm P}e}\gg\omega_{{\rm L}e}$ at the radii considered throughout the paper, the non-linear current is nearly independent of the plasma magnetisation. Hence, we can neglect the effect of the magnetic field on the development of self-modulation.

\section{Self-modulation}
\label{sec:instabilities}

We model the FRB propagation through the circumburst environment by considering a monochromatic linearly-polarised electromagnetic wave that propagates through an electron-ion plasma with constant number density $N$. We are interested in the non-linear effects caused by the finite amplitude of the wave.

The plan of this section is the following. In Section \ref{sec:pump} we find the leading non-linear corrections to the number density and to the transverse velocity of the electrons moving in the field of the electromagnetic wave (called ``pump wave'' below). From the non-linear component of the electron current, we calculate the corrections to the dispersion relation of the pump wave. Using these results, in Section \ref{sec:stability} we study the stability of the pump wave by considering the growth of the two satellite waves ${\bf k}_0\pm {\bf k}$.

\subsection{Electromagnetic pump wave}
\label{sec:pump}

Let $m_i$ and $m_e$ be the ion and the electron mass, and let $e$ and $-e$ be the ion and the electron charge. The transverse electric field of the wave is ${\bf E}={\bf e}_x E_0\cos\chi_0$, where $\chi_0=\omega_0 t-k_0 z$ (the wave propagates along the $z$ axis). We have defined the angular frequency $\omega_0 = 2\pi\nu_0$. We focus on the weakly-relativistic regime $a_0\ll 1$, where $a_0=eE_0/\omega_0 m_ec$ is the strength parameter of the pump wave. We are interested in the case $\omega_0\gg\omega_{{\rm P}e}$, where $\omega_{{\rm P}e}=\sqrt{4\pi Ne^2/m_e}$ is the electron plasma frequency. The ion plasma frequency, $\omega_{{\rm P}i}=\sqrt{4\pi Ne^2/m_i}$, is smaller than the electron plasma frequency by the square root of the mass ratio $m_e/m_i\ll 1$.

\subsubsection{Electron motion in the wave field}

It is useful to introduce the vector potential ${\bf A}$ and the scalar potential $\phi$, so that ${\bf B}=\nabla\times {\bf A}$ and ${\bf E}=-\nabla\phi-(1/c)\pa{\bf A}/\pa t$. Since the electric field of the wave is in the $x$ direction, we have ${\bf A}={\bf e}_x A_x$, where $A_x=-(cE_0/\omega_0)\sin\chi_0=-(m_ec^2/e)a_0\sin\chi_0$. The scalar potential $\phi(\chi_0)$ is calculated below. We work in the Coulomb gauge, $\nabla \cdot {\bf A}=0$.

From the conservation of the generalised momentum, one finds that the $x$ component of the electron momentum is $P_{ex}=eA_x/c=-m_eca_0\sin\chi_0$. At the lowest order in $a_0$, the $x$ component of the electron velocity is therefore $V_{ex}/c=-a_0\sin\chi_0$. The $z$ component of the Lorentz force is $-eV_{ex}B_y/c=(a_0^2/2)\omega_0m_ec\sin(2\chi_0)$, where we have used the approximation $B_y=E_0\cos\chi_0$, which holds in the leading order for the regime of interest, $\omega_0 \gg \omega_{{\rm P}e}$. Neglecting the effect of the electrostatic force, which is justified below, from the $z$ component of the equation of motion one finds that $V_{ez}/c=-(a_0^2/4)\cos (2\chi_0)$.

The electron number density, including the small non-linear corrections, can be calculated from the continuity equation, which gives
\begin{equation}
\label{eq:Ne}
N_e = N\left[1-\frac{1}{4}a_0^2\cos \left(2\chi_0\right)\right] \;,
\end{equation}
where we have used $k_0V_{ez}/\omega_0=V_{ez}/c$ in the leading order. Since the non-linear corrections to the ion number density are of the order of $(m_e^2 / m_i^2)a_0^2\ll a_0^2$, we make the approximation that $N_i=N$. Using the Gauss law, $\pa^2\phi/\pa z^2=4\pi e(N_e-N_i)$, we find that $e\phi/m_ec^2=(a_0^2/16)(\omega_{{\rm P}e}^2/\omega_0^2)\cos(2\chi_0)$. Hence, the electrostatic force is a factor $\omega_{{\rm P}e}^2/\omega_0^2\ll 1$ smaller than the $z$ component of the Lorentz force.

The $x$ component of the electron velocity, including the relativistic corrections of the order of $a_0^2$, is
\begin{equation}
\label{eq:ve}
\frac{V_{ex}}{c}= -a_0\sin\chi_0 \left[1-\frac{1}{4}a_0^2+\frac{1}{4}a_0^2\cos \left(2\chi_0 \right)\right] \;.
\end{equation}
We have used the expansion $V_{ex}/c = P_{ex}/\sqrt{m_e^2c^2+P_{ex}^2}= P_{ex}/m_ec -P_{ex}^3/2m_e^3c^3$, where $P_{ex}/m_ec=-a_0\sin\chi_0$. Since $V_{ez}/c$ is of the order of $a_0^2$, we have neglected its contribution to the electron Lorentz factor.

\subsubsection{Dispersion relation}

The $x$ component of the Amp\`{e}re's law can be presented as
\begin{equation}
\label{eq:wave}
\left(c^2\nabla^2-\frac{\pa^2}{\pa t^2}\right)a=\omega_{{\rm P}e}^2n_e v_{ex} \;,
\end{equation}
where $a=eA_x/m_ec^2=-a_0\sin\chi_0$, $v_{ex}=V_{ex}/c$, and $n_e=N_e/N$. Substituting Eqs. \eqref{eq:Ne}-\eqref{eq:ve} into Eq. \eqref{eq:wave}, we find the dispersion relation
\begin{equation}
\label{eq:DR}
\omega_0^2=c^2k_0^2+\omega_{{\rm P}e}^2-\frac{1}{4}a_0^2\omega_{{\rm P}e}^2 \;.
\end{equation}
The last term is due to the non-linear component of the electron current at the frequency of the pump wave. Eq. \eqref{eq:DR} is consistent with the classical result of \citet[][]{SluijterMontgomery1965} \citep[see also][]{Max+1974}. 

We have neglected the contribution of the linear and of the non-linear components of the ion current to Eqs. \eqref{eq:wave}-\eqref{eq:DR}. These components are much smaller than the corresponding electronic components since the mass ratio is $m_e/m_i\ll 1$.

\subsection{Stability analysis}
\label{sec:stability}

\subsubsection{Wave equation}

Modulations with frequency $\omega$ and wave vector ${\bf k}$ of the pump wave intensity are described by two satellite waves with frequencies $\omega\pm\omega_0$ and wave vectors ${\bf k}\pm k_0{\bf e}_z$. We assume that $k^2\ll k_0^2$, namely the wavelength of the modulations is much longer than the wavelength of the pump wave. Our goal is deriving an equation for the evolution of the satellite waves.

We consider for simplicity perturbations that are independent of $x$. As we discuss below, the dispersion relation is invariant for rotations of the perturbation wave vector, ${\bf k}$, around the $z$ axis. Perturbing Eq. \eqref{eq:wave}, and neglecting terms that are quadratic in the perturbed quantities, we find that
\begin{equation}
\label{eq:deltaa}
\left(c^2\nabla^2-\frac{\pa^2}{\pa t^2}\right)\delta a = \omega_{{\rm P}e}^2\left(v_{ex} \delta n_e + n_e \delta v_{ex} \right)\;.
\end{equation}
We write the perturbed vector potential as $\delta a = \int \delta a(\omega',k_y',k_z')\exp[\i(\omega't-k_y'y-k_z'z)]{\rm d}\omega'{\rm d}k_y'{\rm d}k_z'$, and we introduce analogous definitions for the velocity perturbations and for the density perturbations.\footnote{To avoid heavy notation, we are using the same symbol $\delta a$ for the representation of the vector potential in both coordinate and Fourier space. It will be clear from the context whether we are working in coordinate or in Fourier space.} We substitute these definitions and Eqs. \eqref{eq:Ne}-\eqref{eq:ve} into Eq. \eqref{eq:deltaa}, and we keep terms up to the order of $a_0^2$. As we show in Appendix \ref{sec:derivation}, following this procedure one can derive two coupled equations for the amplitude of the satellite waves, $\delta a_{\pm 1}=\delta a(\omega\pm\omega_0,k_y, k_z\pm k_0)$, namely
\begin{align}
\nonumber
\left[\omega_{+1}^2\right.&-\left. c^2k_{+1}^2 \right]\delta a_{+1} - \omega_{{\rm P}e}^2 \delta v_{ex+1} = \\
\label{eq:deltaa+}
& = \frac{\i}{2}a_0\omega_{{\rm P}e}^2 \left[\delta n_{e0}-\delta n_{e+2}\right] -\frac{1}{8}a_0^2\omega_{{\rm P}e}^2\delta v_{ex-1} \\
\nonumber
\left[\omega_{-1}^2\right.&-\left. c^2k_{-1}^2 \right]\delta a_{-1} - \omega_{{\rm P}e}^2 \delta v_{ex-1} = \\
\label{eq:deltaa-}
& = \frac{\i}{2}a_0\omega_{{\rm P}e}^2 \left[\delta n_{e-2}-\delta n_{e0}\right] - \frac{1}{8}a_0^2\omega_{{\rm P}e}^2 \delta v_{ex+1}
\end{align}
where we have defined $\omega_{\pm 1}=\omega\pm\omega_0$ and ${\bf k}_{\pm 1}={\bf k}\pm k_0{\bf e}_z$. We have also defined $\delta n_{e\pm m}=\delta n_e(\omega\pm m\omega_0,k_y,k_z\pm mk_0)$ and $\delta v_{ex\pm m}=\delta v_{ex}(\omega\pm m\omega_0,k_y,k_z\pm mk_0)$.

Finding the dispersion relation from Eqs. \eqref{eq:deltaa+}-\eqref{eq:deltaa-} is straightforward once the velocity and the density perturbations are expressed as a function of the perturbed vector potential. In Sections \ref{sec:vel} and \ref{sec:dens}, we determine the velocity and the density perturbations. The dispersion relation is then presented in Section \ref{sec:DR}.

\subsubsection{Velocity perturbations}
\label{sec:vel}

Let us define the $x$ component of the electron four-velocity, $u_{ex}=\gamma_e v_{ex}$, where $\gamma_e$ is the electron Lorentz factor. We find that $\delta v_{ex}=\delta u_{ex}/\gamma_e-(u_{ex}/\gamma_e^2)\delta\gamma_e$. Since $u_{ex}^2=\gamma_e^2-1$, we have that $u_{ex}\delta u_{ex}= \gamma_e \delta\gamma_e$ (contributions from $u_{ey}$ and $u_{ez}$ are at least of the order of $a_0^3$). Hence, we find that $\delta v_{ex}=\delta u_{ex}/\gamma_e^3=\delta a/\gamma_e^3$. Using the fact that $\gamma_e=1+(a_0^2/2)\sin^2\chi_0$, we eventually find that
\begin{equation}
\label{eq:deltave}
\delta v_{ex} = \left[1-\frac{3}{4}a_0^2+\frac{3}{4}a_0^2\cos\left(2\chi_0\right)\right]\delta a\;.
\end{equation}
Using the identities presented in Appendix \ref{sec:identities}, from Eq. \eqref{eq:deltave} we find that
\begin{equation}
\label{eq:deltav}
\delta v_{ex\pm 1} = \left[1-\frac{3}{4}a_0^2\right]\delta a_{\pm 1} + \frac{3}{8}a_0^2\delta a_{\mp 1} \;.
\end{equation}
We have neglected terms proportional to $\delta a_{\pm 3}$, which would give corrections of order higher than $a_0^4$ to the dispersion relation.

\subsubsection{Density perturbations}
\label{sec:dens}

Since $\delta n_{e0}$ and $\delta n_{e\pm 2}$ are multiplied by a factor of $a_0$ in Eqs. \eqref{eq:deltaa+} and \eqref{eq:deltaa-}, it is sufficient to calculate exact expressions up to the order of $a_0$. The perturbed continuity equation for the electron fluid is
\begin{equation}
\label{eq:conte}
\frac{\pa\delta n_e}{\pa t} +\nabla\cdot\delta {\bf V}_e=0 \;,
\end{equation}
and the perturbed Euler's equation, which we derive in Appendix \ref{sec:derivation}, is
\begin{align}
\nonumber
\frac{\pa \delta {\bf V}_e}{\pa t} = & -\frac{m_i}{m_e} c_{\rm s}^2\nabla\delta n_e + \frac{e}{m_e}\nabla\delta\phi +\\
\label{eq:eulere}
& +\frac{e}{m_e c}\frac{\pa \delta {\bf A}}{\pa t} -{\bf e}_z c^2k_0a_0\cos\chi_0\delta a + c^2a_0\sin\chi_0\nabla\delta a \;,
\end{align}
where $c_{\rm s}=\sqrt{3k_{\rm B}T/m_i}$ is the thermal velocity of the ions. We have assumed that the electrons and the ions have the same temperature $T$, and that the thermal velocity of the electrons is non-relativistic. Note that the last two terms on the right hand side of Eq. \eqref{eq:eulere} come from the gradient of the perturbed ponderomotive potential, $\delta\phi_{\rm pond}=m_ec^2a\delta a$. As we show in Appendix \ref{sec:derivation}, substituting Eq. \eqref{eq:conte} into the divergence of Eq. \eqref{eq:eulere} and using the perturbed Gauss law, $\nabla^2 \delta \phi=4\pi N e(\delta n_e-\delta n_i)$, one finds that
\begin{align}
\label{eq:deltane0}
& \left[\omega^2-\omega_{{\rm P}e}^2 - \frac{m_i}{m_e}c_{\rm s}^2k^2\right] \delta n_{e0} +\omega_{{\rm P}e}^2\delta n_{i0} = \frac{\i}{2}a_0c^2k^2\left[\delta a_{-1}-\delta a_{+1}\right] \\
\label{eq:deltane+2}
& \left[\omega_{+2}^2-\omega_{{\rm P}e}^2 - \frac{m_i}{m_e}c_{\rm s}^2k_{+2}^2\right] \delta n_{e+2} +\omega_{{\rm P}e}^2\delta n_{i+2} = \frac{\i}{2}a_0c^2k_{+2}^2\delta a_{+1}\\
\label{eq:deltane-2}
& \left[\omega_{-2}^2-\omega_{{\rm P}e}^2 - \frac{m_i}{m_e}c_{\rm s}^2k_{-2}^2\right] \delta n_{e-2} +\omega_{{\rm P}e}^2\delta n_{i-2} = -\frac{\i}{2}a_0c^2k_{-2}^2\delta a_{-1}
\end{align}
where we have defined $\omega_{\pm 2}=\omega\pm 2\omega_0$ and ${\bf k}_{\pm 2}={\bf k}\pm 2k_0{\bf e}_z$. The perturbed continuity equation for the ion fluid is
\begin{equation}
\label{eq:conti}
\frac{\pa\delta n_i}{\pa t} +\nabla\cdot\delta {\bf V}_i=0 \;,
\end{equation}
and the perturbed Euler's equation is
\begin{equation}
\label{eq:euleri}
\frac{\pa \delta {\bf V}_i}{\pa t} = -c_{\rm s}^2\nabla \delta n_i - \frac{e}{m_i}\nabla\delta\phi \;.
\end{equation}
Since the mass ratio is $m_e/m_i\ll 1$, we have neglected the oscillations of the ions in the electromagnetic field of the wave. Substituting Eq. \eqref{eq:conti} into the divergence of Eq. \eqref{eq:euleri}, we find that
\begin{align}
\label{eq:deltani0}
& \left[\omega^2-\omega_{{\rm P}i}^2-c_{\rm s}^2k^2\right] \delta n_{i0} +\omega_{{\rm P}i}^2\delta n_{e0} = 0 \\
\label{eq:deltani+2}
& \left[\omega_{+2}^2-\omega_{{\rm P}i}^2-c_{\rm s}^2k_{+2}^2\right] \delta n_{i+2} +\omega_{{\rm P}i}^2\delta n_{e+2} = 0 \\
\label{eq:deltani-2}
& \left[\omega_{-2}^2-\omega_{{\rm P}i}^2 - c_{\rm s}^2k_{-2}^2\right] \delta n_{i-2} +\omega_{{\rm P}i}^2\delta n_{e-2} = 0
\end{align}
Since $\omega_{{\rm P}i}\ll\omega_0$ and $c_{\rm s}\ll c$, from Eqs. \eqref{eq:deltani+2}-\eqref{eq:deltani-2} one sees that $\delta n_{i\pm 2}=-(\omega_{{\rm P}i}^2/4\omega_0^2)\delta n_{e\pm 2}$, and therefore $\delta n_{i\pm 2}\ll\delta n_{e\pm 2}$.

The low-frequency electron density perturbation $\delta n_{e0}$ is determined by solving Eqs. \eqref{eq:deltane0} and \eqref{eq:deltani0}. We find that
\begin{equation}
\label{eq:deltanl}
\delta n_{e0} = \frac{\i}{2} Q a_0 \left[\delta a_{-1}-\delta a_{+1}\right] \;,
\end{equation}
where
\begin{equation}
\label{eq:Q}
Q= \frac{c^2k^2\left(\omega^2-\omega_{{\rm P}i}^2-c_{\rm s}^2k^2\right)}{\omega^2\left(\omega^2-\omega_{{\rm P}e}^2\right) - \frac{m_i}{m_e}c_{\rm s}^2k^2 \left(\omega^2-2\omega_{{\rm P}i}^2-c_{\rm s}^2k^2\right)} \;.
\end{equation}
Since $\delta n_{i\pm 2}\ll\delta n_{e\pm 2}$, the high-frequency electron density perturbations $\delta n_{e\pm 2}$ are simply determined by solving Eqs. \eqref{eq:deltane+2} and \eqref{eq:deltane-2}. We find that
\begin{equation}
\label{eq:deltanh}
\delta n_{e\pm 2} = \pm\frac{\i}{2}a_0\delta a_{\pm 1} \;,
\end{equation}
where we have used the fact that $(m_i/m_e)c_{\rm s}^2\ll c^2$ since the thermal velocity of the electrons is non-relativistic.

An important point is the following. In the general case when the perturbations depend also on $x$, one should consider the contribution of the perturbed electrostatic potential to Eq. \eqref{eq:deltaa}, and the contributions of the perturbed electrostatic potential, of the perturbed ponderomotive potential, and of the perturbed density gradient to Eq. \eqref{eq:deltave}. The non-vanishing Fourier components of these additional terms have the frequencies $\omega$ and $\omega\pm 2\omega_0$. Hence, Eqs. \eqref{eq:deltaa+}, \eqref{eq:deltaa-}, and \eqref{eq:deltav} remain the same since they describe Fourier components at the frequency $\omega\pm \omega_0$. One therefore sees that the dispersion relation, Eq. \eqref{eq:DRfinal}, is invariant for rotations of the perturbation wave vector, ${\bf k}$, around the $z$ axis.

\subsubsection{Dispersion relation}
\label{sec:DR}

The procedure to obtain the dispersion relation is the following. We substitute Eqs. \eqref{eq:deltav}, \eqref{eq:deltanl}, and \eqref{eq:deltanh} into Eqs. \eqref{eq:deltaa+}-\eqref{eq:deltaa-}. We obtain a linear homogeneous system of two equations for $\delta a_{\pm 1}$. The dispersion relation is found by imposing the condition that the determinant of the matrix of the coefficients vanishes. Using the fact that $\omega_{\pm 1}^2-c^2k_{\pm 1}^2 = (\omega^2-c^2k^2) \pm 2(\omega_0\omega-c^2k_0k_z) +\omega_{{\rm P}e}^2(1-a_0^2/4)$, which can be obtained using Eq. \eqref{eq:DR}, we find that
\begin{align}
\nonumber
\left(\omega^2 - c^2k^2\right)^2 & -4\left(\omega_0\omega-c^2k_0k_z\right)^2 + \\
\label{eq:DRfinal}
& +\frac{1}{2}a_0^2\omega_{{\rm P}e}^2\left(1-Q\right)\left(\omega^2-c^2 k^2\right) =0 \;,
\end{align}
where $Q$ is defined in Eq. \eqref{eq:Q}. The general dispersion relation, Eq. \eqref{eq:DRfinal}, is cumbersome due to the complicated dependence of $Q$ on the parameters of the problem. Hence, it is convenient to discuss the relevant regimes separately, which we do in the following.

When $\omega^2\ll\omega_{{\rm P}i}^2$ and $c_{\rm s}^2k^2\ll\omega^2$, one finds that $Q=(\omega_{{\rm P}i}^2/\omega_{{\rm P}e}^2)(c^2k^2/\omega^2)$, and the dispersion relation is
\begin{align}
\nonumber
\left(\omega^2\right. & - \left. c^2k^2\right)^2 -4\left(\omega_0\omega-c^2k_0k_z\right)^2 + \\
\label{eq:DR1}
& +\frac{1}{2}a_0^2\left(\omega_{{\rm P}e}^2-\frac{c^2k^2}{\omega^2}\omega_{{\rm P}i}^2\right)\left(\omega^2-c^2 k^2\right) =0 \;.
\end{align}
When $\omega^2\ll\omega_{{\rm P}i}^2$ and $\omega^2\ll c_{\rm s}^2k^2\ll\omega_{{\rm P}i}^2$, one finds that $Q=-(1/2)(\omega_{{\rm P}i}^2/\omega_{{\rm P}e}^2)(c^2/c_{\rm s}^2)$, and the dispersion relation is
\begin{align}
\nonumber
\left(\omega^2\right. & - \left. c^2k^2\right)^2 -4\left(\omega_0\omega-c^2k_0k_z\right)^2 + \\
\label{eq:DR2}
& +\frac{1}{2}a_0^2\left(\omega_{{\rm P}e}^2+\frac{1}{2}\frac{c^2}{c_{\rm s}^2}\omega_{{\rm P}i}^2\right)\left(\omega^2-c^2 k^2\right) =0 \;.
\end{align}
When $\omega_{{\rm P}i}^2\ll\omega^2\ll\omega_{{\rm P}e}^2$ and $c_{\rm s}^2k^2\ll\omega_{{\rm P}i}^2$, one finds that $Q=-c^2k^2/\omega_{{\rm P}e}^2$, and the dispersion relation is
\begin{align}
\nonumber
\left(\omega^2\right. & - \left. c^2k^2\right)^2 -4\left(\omega_0\omega-c^2k_0k_z\right)^2 + \\
\label{eq:DR3}
& +\frac{1}{2}a_0^2\left(\omega_{{\rm P}e}^2+c^2k^2\right)\left(\omega^2-c^2 k^2\right) =0 \;.
\end{align}
When $\omega_{{\rm P}e}^2\ll\omega^2$ and $(\omega_{{\rm P}e}^2/\omega_{{\rm P}i}^2)c_{\rm s}^2k^2\ll\omega^2$, one finds that $Q=c^2k^2/\omega^2$, and the dispersion relation is
\begin{align}
\nonumber
\left(\omega^2\right. & - \left. c^2k^2\right)^2 -4\left(\omega_0\omega-c^2k_0k_z\right)^2 + \\
\label{eq:DR4}
& +\frac{1}{2}a_0^2\omega_{{\rm P}e}^2\left(1-\frac{c^2k^2}{\omega^2}\right)\left(\omega^2-c^2 k^2\right) =0 \;.
\end{align}
When $\omega_{{\rm P}e}^2\ll\omega^2$ and $\omega^2\ll (\omega_{{\rm P}e}^2/\omega_{{\rm P}i}^2) c_{\rm s}^2k^2$, or when $\omega^2\ll\omega_{{\rm P}e}^2$ and $\omega_{{\rm P}i}^2\ll c_{\rm s}^2k^2$, one finds that $Q=-(\omega_{{\rm P}i}^2/\omega_{{\rm P}e}^2)(c^2/c_{\rm s}^2)$, and the dispersion relation is
\begin{align}
\nonumber
\left(\omega^2\right. &- \left. c^2k^2\right)^2 -4\left(\omega_0\omega-c^2k_0k_z\right)^2 + \\
\label{eq:DR5}
& +\frac{1}{2}a_0^2\left(\omega_{{\rm P}e}^2+\frac{c^2}{c_{\rm s}^2}\omega_{{\rm P}i}^2\right)\left(\omega^2-c^2 k^2\right) =0 \;.
\end{align}
In the following we characterise the unstable modes in the different regimes.

\subsubsection{Unstable modes}

In order to characterise the most unstable modes, it is convenient to define $\omega= c^2k_0k_z/\omega_0+\Delta\omega$. With this definition, we have $(\omega_0\omega-c^2k_0k_z)^2 = \omega_0^2(\Delta\omega)^2$. One can also make the approximation that $\omega^2 - c^2k^2= -c^2k_y^2 -c^2k_z^2\omega_{{\rm P}e}^2/\omega_0^2$. The reason is that both $(\Delta\omega)^2$ and $ck_z(\Delta\omega)$ are much smaller than $c^2k_y^2$, which can be verified a posteriori case by case (see Tables \ref{table1} and \ref{table2}). Finally, since we will find that $\Delta\omega$ is purely imaginary for the unstable modes, the instability is purely growing in the frame moving with the group velocity of the pump wave.

It is convenient to start considering the modes that are not affected by the ion dynamics and by the thermal motions. When $(c^2k^2/\omega^2)\omega_{{\rm P}i}^2\ll \omega_{{\rm P}e}^2$, Eq. \eqref{eq:DR1} gives
\begin{equation}
\label{eq:sm}
4\omega_0^2\left(\Delta\omega\right)^2 = \left(c^2k_y^2+\frac{\omega_{{\rm P}e}^2}{\omega_0^2}c^2k_z^2\right) \left(c^2k_y^2+\frac{\omega_{{\rm P}e}^2}{\omega_0^2}c^2k_z^2-\frac{1}{2}a_0^2\omega_{{\rm P}e}^2\right) \;,
\end{equation}
which is consistent with the results of \citet[][]{Max+1974}. There are two important effects that determine the behaviour of the modes, namely (i) the non-linear component of the current, which gives the destabilising contribution proportional to $a_0^2$ to the dispersion relation; (ii) diffraction, which stabilises the modes with a short wavelength by softening the gradients of the radiation intensity.

From Eq. \eqref{eq:sm}, the maximum growth rate of the instability is found when $c^2k_y^2+(\omega_{{\rm P}e}^2/\omega_0^2)c^2k_z^2=a_0^2\omega_{{\rm P}e}^2/4$, which gives $(\Delta\omega)^2 = -a_0^4\omega_{{\rm P}e}^4/64\omega_0^2$. Since typically $ck_y\simeq a_0\omega_{{\rm P}e}/2$ and $ck_z\simeq a_0\omega_0/2$, the most unstable modes are elongated in the direction perpendicular to the direction of propagation of the pump wave. We neglect the effect of the unstable modes with a wave vector significantly different than the typical one, since these modes occupy a small volume of the phase space. Since $(c^2k^2/\omega^2)\omega_{{\rm P}i}^2\simeq \omega_{{\rm P}i}^2\ll\omega_{{\rm P}e}^2$, neglecting the effect of the ion motion is justified.

Eq. \eqref{eq:DR1} can be used when $c^2k_z^2\simeq\omega^2\ll\omega_{{\rm P}i}^2$, which requires that $a_0\omega_0\ll\omega_{{\rm P}i}$. When instead $\omega_{{\rm P}i}\ll a_0\omega_0\ll\omega_{{\rm P}e}$, one should use Eq. \eqref{eq:DR3}. Using the fact that $c^2k^2\simeq a_0^2\omega_0^2\ll\omega_{{\rm P}e}^2$, Eq. \eqref{eq:DR3} gives the same dispersion relation as before, Eq. \eqref{eq:sm}.

Finally, from Eq. \eqref{eq:DR4} one sees that self-modulations are stabilised when $\omega_{{\rm P}e}^2\ll c^2k_z^2\simeq\omega^2$. Hence, when $\omega_{{\rm P}e}\ll a_0\omega_0$ the most unstable modes have the same transverse wave number as before, $ck_y\simeq a_0\omega_{{\rm P}e}/2$, while the longitudinal wave number is $ck_z\lesssim \omega_{{\rm P}e}$. The growth rate remains $(\Delta\omega)^2=-a_0^4\omega_{{\rm P}e}^4/64\omega_0^2$.

The effect of the thermal motions can be always neglected since $(\omega_{{\rm P}e}^2/\omega_{{\rm P}i}^2)c_{\rm s}^2k^2\ll\omega^2$. This is the case because,  if the thermal velocity of the electrons is non-relativistic, one finds that $(\omega_{{\rm P}e}^2/\omega_{{\rm P}i}^2)c_{\rm s}^2k^2\simeq (\omega_{{\rm P}e}^2/\omega_{{\rm P}i}^2)c_{\rm s}^2k_z^2\ll c^2k_z^2\simeq\omega^2$. Hence, we do not need to discuss Eqs. \eqref{eq:DR2} and \eqref{eq:DR5}.

We conclude that there is a first class of unstable modes that are independent of the ion dynamics and of the thermal motions \citep[see also][]{Max+1974}. For these modes, we may estimate the most unstable wave number as
\begin{align}
\label{eq:k1y}
k_y & \simeq a_0\frac{\omega_{{\rm P}e}}{c} \\
\label{eq:k1z}
k_z & \simeq \min\left[a_0\frac{\omega_0}{c},\; \frac{\omega_{{\rm P}e}}{c} \right]
\end{align}
and the growth rate as
\begin{equation}
\label{eq:g1}
\Gamma \simeq a_0^2\frac{\omega_{{\rm P}e}^2}{\omega_0} \;.
\end{equation}
These results are summarised in Table \ref{table1}. Since $k_y\ll k_z$, the modulations are elongated in the direction perpendicular to the direction of propagation of the electromagnetic pump wave.

\begin{table}
\begin{center}
\begin{tabular}{| c | c | c | c |}
\hline
$ck_y$ & $ck_z$ & $\Gamma$ & range of $a_0$ \\ [0.5ex] 
\hline\hline
$a_0\omega_{{\rm P}e}$ & $a_0\omega_0$ & $a_0^2 \omega_{{\rm P}e}^2 / \omega_0$ & $a_0\lesssim \omega_{{\rm P}e}/\omega_0$ \\ [0.5ex] 
\hline
$a_0\omega_{{\rm P}e}$ & $\omega_{{\rm P}e}$ & $a_0^2 \omega_{{\rm P}e}^2 / \omega_0$ & $a_0\gtrsim \omega_{{\rm P}e}/\omega_0$ \\ [0.5ex] 
\hline
\end{tabular}
\caption{\label{table1} Wave number in the transverse direction ($k_y$) and in the longitudinal direction ($k_z$), and growth rate ($\Gamma$) of the unstable modes that are independent of the ion dynamics and of the thermal motions (see also Eqs. \ref{eq:k1y}-\ref{eq:g1}). For these modes, one finds that $k_y\ll k_z$, i.e. the modulations are elongated in the direction perpendicular to the direction of propagation of the electromagnetic pump wave.}
\end{center}
\end{table}

\begin{table}
\begin{center}
\begin{tabular}{| c | c | c | c |}
\hline
$ck_y$ & $ck_z$ & $\Gamma$ & range of $a_0$ \\ [0.5ex] 
\hline\hline
$a_0\beta_{\rm s}^{-1}\omega_{{\rm P}i}$ & $a_0\omega_{{\rm P}i}$ & $a_0^2 \beta_{\rm s}^{-2} \omega_{{\rm P}i}^2 / \omega_0$ & $a_0\lesssim \beta_{\rm s}^2 \omega_0/\omega_{{\rm P}i}$ \\ [0.5ex] 
\hline
$\sqrt{a_0\omega_0\omega_{{\rm P}i}}$ & $a_0\omega_{{\rm P}i}$ & $a_0\omega_{{\rm P}i}$ & $a_0\gtrsim \beta_{\rm s}^2 \omega_0/\omega_{{\rm P}i}$ \\ [0.5ex] 
\hline
\end{tabular}
\caption{\label{table2} Wave number in the transverse direction ($k_y$) and in the longitudinal direction ($k_z$), and growth rate ($\Gamma$) of the unstable modes that depend on the ion dynamics and on the thermal motions (see also Eqs. \ref{eq:k2y}-\ref{eq:g2}). We have defined $\beta_{\rm s}=c_{\rm s}/c$, where $c_{\rm s}$ is the thermal velocity of the ions. For these modes, one finds that $k_y\gg k_z$, i.e. the modulations are elongated in the direction of propagation of the electromagnetic pump wave.}
\end{center}
\end{table}

We now consider the modes where the effect of the ion dynamics and of the thermal motions is important. When $(c^2k^2/\omega^2)\omega_{{\rm P}i}^2\gg \omega_{{\rm P}e}^2$ and $(\Delta\omega)^2\gg c^2k_z^2$, one may approximate $\omega^2 - c^2k^2= -c^2k_y^2$ and $(c^2k^2/\omega^2)\omega_{{\rm P}i}^2=(c^2k_y^2/(\Delta\omega)^2)\omega_{{\rm P}i}^2$. Hence, Eq. \eqref{eq:DR1} gives
\begin{equation}
\label{eq:sf1}
4\omega_0^2\left(\frac{\Delta\omega}{ck_y}\right)^4 -c^2k_y^2\left(\frac{\Delta\omega}{ck_y}\right)^2-\frac{1}{2}a_0^2\omega_{{\rm P}i}^2 =0 \;,
\end{equation}
which is consistent with the results of \citet[][]{Drake+1974}. According to Eq. \eqref{eq:sf1}, the wave number of the most unstable modes is $ck_y \gg \sqrt{a_0\omega_0\omega_{{\rm P}i}}$, and the corresponding growth rate is $ (\Delta\omega)^2 = -a_0^2\omega_{{\rm P}i}^2/2$. The condition that $(\Delta\omega)^2\gg c^2k_z^2$ gives $ck_z\ll a_0\omega_{{\rm P}i}$. When instead $(c^2k^2/\omega^2)\omega_{{\rm P}i}^2\gg \omega_{{\rm P}e}^2$ and $(\Delta\omega)^2\ll c^2k_z^2$, one may approximate $(c^2k^2/\omega^2)\omega_{{\rm P}i}^2=(k_y^2/k_z^2)\omega_{{\rm P}i}^2$, in which case Eq. \eqref{eq:DR1} does not give any instability.

Eq. \eqref{eq:DR1} can be used when $c_{\rm s}^2k^2\ll\omega^2$, which requires that $a_0\gg (c_{\rm s}^2/c^2)(\omega_0/\omega_{{\rm P}i})$. When instead $\omega^2\ll c_{\rm s}^2k^2$, one should use Eq. \eqref{eq:DR2}. Using the fact that $(\omega_{{\rm P}e}^2/\omega_{{\rm P}i}^2)c_{\rm s}^2\ll c^2$ since the thermal velocity of the electrons is non-relativistic, we find that
\begin{equation}
\label{eq:sf2}
4\omega_0^2\left(\Delta \omega\right)^2= c^2k_y^2 \left(c^2k_y^2-\frac{1}{4}a_0^2\frac{c^2}{c_{\rm s}^2}\omega_{{\rm P}i}^2\right) \;,
\end{equation}
which is consistent with the results of \citet[][]{Drake+1974}. According to Eq. \eqref{eq:sf2}, the wave number of the most unstable mode is $ck_y = (1/2\sqrt{2})a_0(c/c_{\rm s})\omega_{{\rm P}i}$, and the corresponding growth rate is $ (\Delta\omega)^2 = -(a_0^4/256)(c^4/c_{\rm s}^4)(\omega_{{\rm P}i}^4/\omega_0^2)$. The condition that $\omega^2\ll c_{\rm s}^2k^2$ requires that $ck_z\ll c_{\rm s}k_y$, which gives $ck_z\ll a_0\omega_{{\rm P}i}$. Finally, it turns out that the regime where Eq. \eqref{eq:DR5} is valid is not relevant for self-modulation.

We conclude that there is a second class of unstable modes that depend on the ion dynamics and on the thermal motions \citep[see also][]{Drake+1974}. For these modes, we may estimate the most unstable wave number as
\begin{align}
\label{eq:k2y}
k_y & \simeq \min\left[ a_0\frac{\omega_{{\rm P}i}}{c_{\rm s}},\; \frac{\sqrt{a_0\omega_0\omega_{{\rm P}i}}}{c} \right] \\
\label{eq:k2z}
k_z & \simeq a_0\frac{\omega_{{\rm P}i}}{c}
\end{align}
and the growth rate as
\begin{equation}
\label{eq:g2}
\Gamma \simeq \min\left[ a_0^2\frac{c^2}{c_{\rm s}^2}\frac{\omega_{{\rm P}i}^2}{\omega_0} ,\; a_0\omega_{{\rm P}i} \right] \;.
\end{equation}
These results are summarised in Table \ref{table2}. Since $k_y\gg k_z$, the modulations are elongated in the direction of propagation of the electromagnetic pump wave.

\section{Implications for Fast Radio Bursts}
\label{sec:FRBs}

\begin{figure*}{\vspace{3mm}} 
\centering
\includegraphics[width=0.99\textwidth]{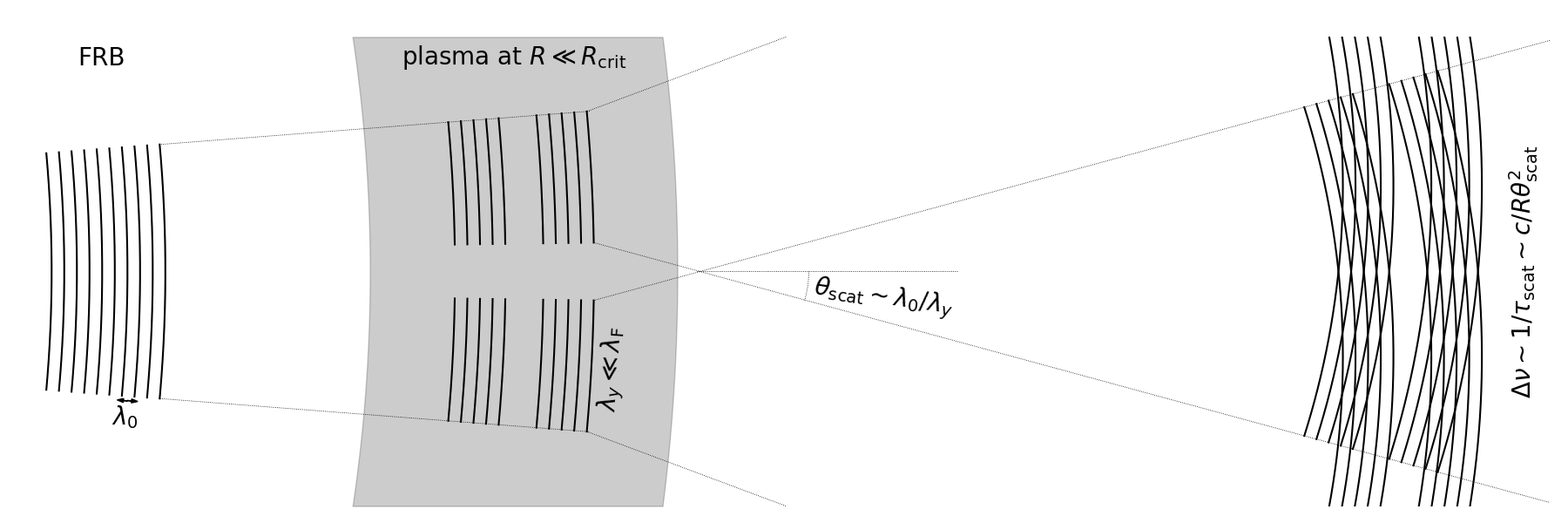}
\caption{Sketch (not to scale) of the physical scenario discussed in Section \ref{sec:freq}. The FRB electromagnetic wave (black lines) interacts with a plasma slab (grey region) located at the distance $R\ll R_{\rm crit}$ from the center, where $R_{\rm crit}$ is given by Eq. \eqref{eq:Rcrit}. Self-modulation breaks the burst into pancakes whose transverse size, $\lambda_y$, is much smaller than the Fresnel scale, $\lambda_{\rm F}=\sqrt{\lambda_0R}$, where $\lambda_0$ is the wavelength of the electromagnetic wave. Since $\lambda_y\ll\lambda_{\rm F}$, diffraction broadens the angular size of the pancakes by $\theta_{\rm scat}\sim\lambda_0/\lambda_y\gg\lambda_y/R$, and the observer sees the interference pattern of a large number of pancakes. The typical scattering time, $\tau_{\rm scat}\sim R\theta_{\rm scat}^2/c\sim R\lambda_0^2/c\lambda_y^2 \gtrsim \lambda_0/c$, corresponds to a frequency modulation with a large bandwidth, $\Delta\nu\sim 1/\tau_{\rm scat}\lesssim{\rm GHz}$ (see Eq. \ref{eq:deltaomega1}).
}
\label{fig:1}
\end{figure*}

\begin{figure*}{\vspace{3mm}} 
\centering
\includegraphics[width=0.99\textwidth]{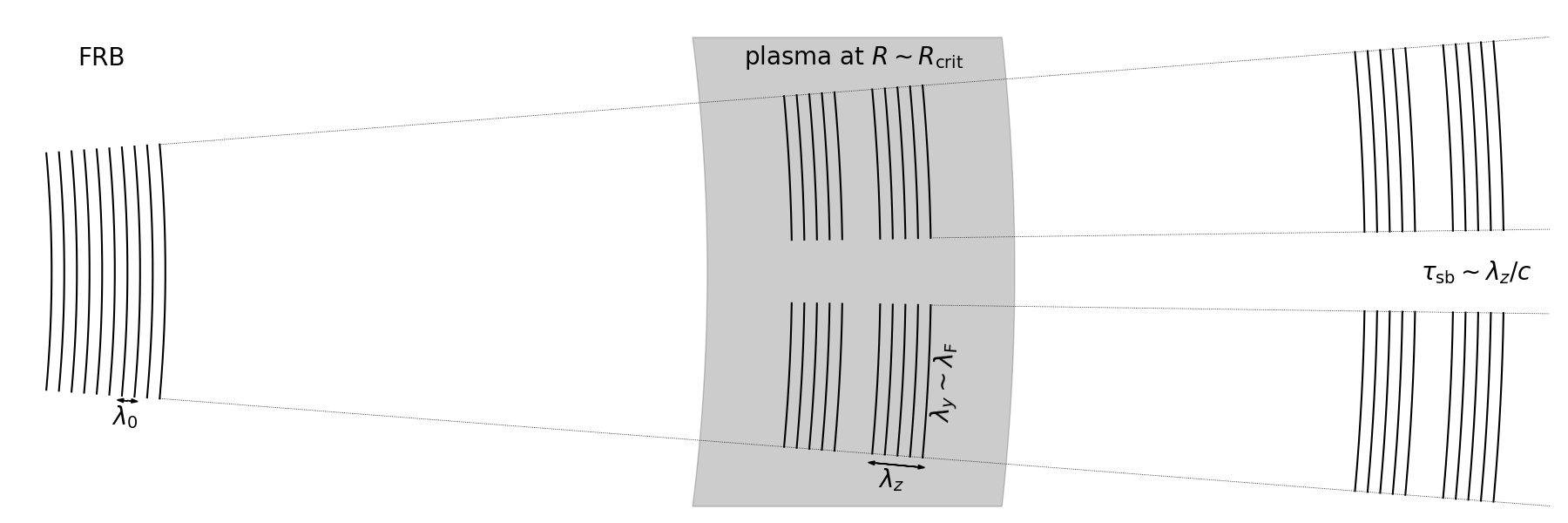}
\caption{Sketch (not to scale) of the physical scenario discussed in Section \ref{sec:t}. The FRB electromagnetic wave (black lines) interacts with a plasma slab (grey region) located at the distance $R\sim R_{\rm crit}$ from the center, where $R_{\rm crit}$ is given by Eq. \eqref{eq:Rcrit}. Self-modulation breaks the burst into pancakes whose transverse size, $\lambda_y$, is comparable to the Fresnel scale, $\lambda_{\rm F}=\sqrt{\lambda_0R}$. Since $\lambda_y\sim\lambda_{\rm F}$, the effect of diffraction is weaker. The observer receives sub-bursts with a typical duration $\tau_{\rm sb}\sim\lambda_z/c$, where $\lambda_z$ is the radial width of the pancakes. One typically finds that $\tau_{\rm sb}\sim 10{\rm\;\mu s}$ (see Eqs. \ref{eq:tm1} and \ref{eq:tm2}).
}
\label{fig:2}
\end{figure*}

In this Section we discuss the observational signatures of the modes that are independent of the ion dynamics and of the thermal motions. These modes, which are described by Eqs. \eqref{eq:k1y}-\eqref{eq:g1}, have a typical wave number $k_y\simeq a_0\omega_{{\rm P}e}/c$ in the transverse direction and $k_z\simeq \min[a_0\omega_0/c,\; \omega_{{\rm P}e}/c]$ in the longitudinal direction (the direction of the pump wave propagation). The growth rate of the modulations is $\Gamma\simeq a_0^2\omega_{{\rm P}e}^2/\omega_0$. Self-modulation saturates when the modulation amplitude becomes comparable to unity, so that the wave packet breaks up into pancakes of transverse size $\lambda_y = 2\pi/k_y$ and radial width $\lambda_z = 2\pi/k_z\ll \lambda_y$. The exact shape of these pancakes may depend on the form of the seed perturbations amplified by the instability. The characteristic separation between the pancakes should be comparable to their sizes.

We consider an idealised setup where the burst radiation interacts with a uniform plasma slab located at the distance $R$ from the center. We assume the thickness of the plasma slab to be slightly smaller than $R$, so that the geometry of the problem is essentially planar inside the slab. Let the burst have a center frequency $\nu_0=\omega_0/2\pi$, and a bandwidth that is comparable with $\nu_0$. Since self-modulation is purely growing in the frame moving with the group velocity of the wave, the instability may develop if the timescale for the instability to grow, $t_{\rm growth}\sim 10/\Gamma$, is shorter than the wave crossing time of the slab, $t_{\rm cross}\sim R/c$. We have taken into account that $\gtrsim 10$ $e$-folding times are needed for the instability to grow from a seed perturbation. Using Eq. \eqref{eq:a0} to express $a_0$, the condition that $t_{\rm growth}\lesssim t_{\rm cross}$ gives $R\lesssim R_{\rm crit}$, where
\begin{equation}
\label{eq:Rcrit}
R_{\rm crit}\sim 0.03 \left(\frac{S_{\nu_0}}{{\rm Jy}}\right) \left(\frac{\nu_0}{{\rm GHz}}\right)^{-2} \left(\frac{N}{10^2{\rm\; cm}^{-3}}\right) \left(\frac{D}{{\rm Gpc}}\right)^2 {\rm\; pc}\;.
\end{equation}
Hence, the modes described by Eqs. \eqref{eq:k1y}-\eqref{eq:g1} may become unstable even at relatively large distances from the source.\footnote{Note that the effect of induced Compton and Raman scattering, which may be important close to the source, is negligible at parsec distances \citep[e.g.][]{Lyubarsky2008}. The reason is that the efficiency of these processes is limited by the short duration of the burst.} One may also express $R_{\rm crit}$ in terms of the isotropic equivalent of the burst luminosity, $L\sim 4\pi D^2S_{\nu_0}\nu_0$, which gives $R_{\rm crit}\sim 0.03 (\nu_0/{\rm GHz})^{-3}(N/10^2{\rm\; cm}^{-3})(L/10^{42}{\rm erg\; s}^{-1}){\rm\; pc}$.

As the burst exits the plasma slab, the evolution of the pancakes that have formed due to the instability is determined by (i) the spherical expansion of the wave front; (ii) the effect of diffraction. Initially, the angular size of the pancakes is $\theta_{\rm sph}\simeq \lambda_y/R$. Individual pancakes are diffracted similar to light passing through a circular aperture of radius $\sim\lambda_y$. Effectively, diffraction broadens the angular size of the pancake by $\theta_{\rm scat}\sim k_y/k_0\sim \lambda_0/\lambda_y$, where $\lambda_0= 2\pi/k_0$ is the wavelength of the pump wave. Since we are assuming the angular separation between the pancakes to be comparable with $\theta_{\rm sph}$, the pancakes interfere with each other if $\theta_{\rm scat}\gtrsim\theta_{\rm sph}$ (see Figure \ref{fig:1}), while the effect of interference becomes negligible if $\theta_{\rm scat}\lesssim\theta_{\rm sph}$ (see Figure \ref{fig:2}). The condition that $\theta_{\rm scat}\lesssim\theta_{\rm sph}$ may be presented as $\lambda_y \gtrsim \lambda_{\rm F}$, where $\lambda_{\rm F}=\sqrt{\lambda_0 R}$ is the Fresnel scale. As we show in Sections \ref{sec:freq} and \ref{sec:t}, the ratio $\lambda_y /\lambda_{\rm F}$ (and therefore the observational signatures of self-modulation) is determined by the position of the plasma slab.

The observer sees a patch of the plasma slab of radius $R\theta_{\rm scat}$, and the corresponding angular size is $(R/D)\theta_{\rm scat}$. For our fiducial parameter choice, one finds that $\theta_{\rm scat}\sim 2\times 10^{-8}(R_{\rm crit}/R) {\rm\; rad}\sim 5 (R_{\rm crit}/R) {\rm\; mas}$, and therefore $(R/D)\theta_{\rm scat}\sim 2\times 10^{-10}{\rm\; mas}$. Hence, there is not any significant broadening of the source image.

In analogy with the standard results of pulsar scintillation theory \citep[e.g.][]{Narayan1992}, one would expect the effects of interference to disappear if the source size, $R_{\rm s}$, exceeds the transverse size of the pancakes, $\lambda_y$. The condition that $R_{\rm s}\lesssim \lambda_y$ gives an upper limit on the source size, $R_{\rm s}\lesssim 10^9(R/R_{\rm crit}){\rm\; cm}$, which can be satisfied by millisecond duration bursts.\footnote{In the synchrotron maser emission model of FRBs \citep[e.g.][]{Beloborodov2017, Beloborodov2020, Metzger+2019}, the burst is emitted by a relativistic blast wave propagating with Lorentz factor $\Gamma_{\rm sh}$. Then radiation is Doppler collimated within an angle of $1/\Gamma_{\rm sh}$, and the effective source size is $R_{\rm s}\sim R_{\rm em}/\Gamma_{\rm sh}$, where $R_{\rm em}$ is the emission radius.}

In addition, the effect of turbulence in the circumburst medium can be neglected if the density fluctuations on scales smaller than the observable patch, $r \ll R\theta_{\rm scat}$, produce small phase perturbations, $\Delta\phi\lesssim\pi$. In order to estimate $\Delta\phi$, we follow the classical approach of \citet[][]{Scheuer1968}. Assuming a Kolmogorov-like spectrum, the density fluctuations on a scale $r$ are $\Delta N\sim sN(r/R)^{1/3}$, where $s$ is a numerical factor quantifying the turbulence amplitude. Due to the fluctuation of the refraction index, $\Delta n\sim (e^2\lambda_0^2/2\pi m_ec^2)\Delta N$, the phase is perturbed by $\delta\phi\sim (r/\lambda_0)\Delta n$ while the wave propagates over a distance $r$. As the wave crosses a distance $R$, the contribution of the turbulent eddies with size $r$ to the random walk of phase is $\Delta\phi\sim (R/r)^{1/2}\delta\phi\sim (s/2\pi)(e^2/m_ec^2)\lambda_0NR(r/R)^{5/6}$. The condition that $\Delta\phi\lesssim\pi$ gives $s\lesssim 0.7(R\theta_{\rm scat}/r)^{5/6}(R_{\rm crit}/R)^{1/6}$. Since we are interested in scales $r\ll R\theta_{\rm scat}$, this condition can be satisfied even for a strong turbulence, e.g. with $s\sim 1$.

The observational signatures of the modes that depend on the ion dynamics and on the thermal motions, which are described by Eqs. \eqref{eq:k2y}-\eqref{eq:g2}, are discussed in Appendix \ref{sec:scattering}. These modes may give an important contribution to the scattering time of FRBs. However, these modes can only develop very close to the source (we find that $R_{\rm crit}\sim 2\times 10^{-5}{\rm\; pc}$), since at larger distances the radial width of the pancakes would exceed the length of the burst itself. Since the properties of the plasma are poorly constrained at these small radii, the results of Appendix \ref{sec:scattering} are very speculative.

\subsection{Frequency structure}
\label{sec:freq}

The physical scenario discussed in this section is sketched in Figure \ref{fig:1}. We consider the effect of a plasma slab at $R\ll R_{\rm crit}$, which corresponds to $\Gamma R/c\gg 10$. Using the fact that $\Gamma\simeq c^2k_y^2/\omega_0$, one sees that $\lambda_y\ll\sqrt{\lambda_0 R}$. Hence, the transverse size of the pancakes is much smaller than the Fresnel scale, and the observer sees the interference pattern of a large number of pancakes. The typical scattering time is $\tau_{\rm scat}\simeq R\theta_{\rm scat}^2/c$, where $\theta_{\rm scat}\simeq \lambda_0/\lambda_y$, and the corresponding frequency modulation bandwidth, $\Delta\nu\simeq 1/\tau_{\rm scat}$, is
\begin{equation}
\label{eq:deltaomega1}
\Delta\nu\sim 0.6\left(\frac{\nu_0}{{\rm GHz}}\right) \left(\frac{R}{R_{\rm crit}}\right) {\rm\; GHz} \;.
\end{equation}
Hence, we expect $\Delta\nu$ to be smaller than $\nu_0$. If the plasma is confined into a thin slab, using the definition of $R_{\rm crit}$, Eq. \eqref{eq:Rcrit}, we find that $\Delta\nu\propto\nu_0^3$. The dependence of $\Delta\nu$ on $\nu_0$ is less clear if there is a continuous distribution of plasma along the line of sight. In this case, modulation may occur in a wide range of frequency bands, which corresponds to the wide distribution of $R/R_{\rm crit}$. If the value of $R/R_{\rm crit}$ giving the dominant contribution to $\Delta\nu$ were independent of $\nu_0$, one would find that $\Delta\nu\propto\nu_0$. In general, we expect the frequency modulation bandwidth, $\Delta\nu$, to increase with the center frequency of the burst, $\nu_0$.

The study of high-signal-to-noise bursts from the repeating FRB 121102 has shown that the bursts have a complex time-frequency structure, which includes sub-bursts with a finite duration and bandwidth \citep[][]{Hessels+2019}. The observed bandwidth is $\sim 100-400{\rm\; MHz}$ for the bursts with a frequency of $1.4$ and $2.0{\rm\; GHz}$, and $\sim 1{\rm\; GHz}$ for the bursts with a frequency of $6.5{\rm\; GHz}$, which is consistent with the trend expected from Eq. \eqref{eq:deltaomega1}.\footnote{Interestingly, the high-frequency interpulse of the Crab pulsar also shows a banded frequency structure with $\Delta\nu\propto\nu_0$ \citep[e.g.][]{HankinsEilek2007, Hankins+2016}. However, in this case the instability could only develop well inside the radius of the pulsar wind termination shock, because the pulsar radio waves are weak compared with FRBs. Self-modulation in a magnetised pair plasma such as the Crab pulsar wind is an interesting topic for future investigation.} Bright FRBs in the ASKAP sample may also show some broadband frequency structure \citep[e.g.][]{Shannon+2018}.

\subsection{Time structure}
\label{sec:t}

The physical scenario discussed in this section is sketched in Figure \ref{fig:2}. We consider the effect of a plasma slab at $R\sim R_{\rm crit}$, in which case $\Gamma R/c \sim 10$ and $\lambda_y\sim \sqrt{\lambda_0 R}$. Now the transverse size of the pancakes is comparable with the Fresnel scale, and hence the broadening of the pancakes exiting the plasma slab is marginal.\footnote{Self-modulation cannot reduce the opening angle of the FRB emission since the angular scale of the pancakes, $\lambda_y/R\sim 2\times 10^{-8}{\rm\; rad}$, is much smaller than the opening angle. It is therefore unlikely that the event rate of FRBs is underestimated due to non-linear propagation effects, as recently proposed by \citet[][]{YangZhang2020}.} In this regime, the observer receives sub-burst of duration $\tau_{\rm sb}\sim \lambda_z/c$, where $\lambda_z=2\pi/k_z$ is the radial width of the pancakes. One finds that $\tau_{\rm sb}$ is the longest between
\begin{equation}
\label{eq:tm1}
\tau_{\rm sb}\sim 10 \left(\frac{N}{10^2{\rm\; cm}^{-3}}\right)^{-1/2}{\rm\; \mu s}
\end{equation}
and
\begin{equation}
\label{eq:tm2}
\tau_{\rm sb}\sim 4 \left(\frac{S_{\nu_0}}{{\rm Jy}}\right)^{1/2} \left(\frac{\nu_0}{{\rm GHz}}\right)^{-5/2} \left(\frac{N}{10^2{\rm\; cm}^{-3}}\right) \left(\frac{D}{{\rm Gpc}}\right) {\rm\; \mu s} \;,
\end{equation}
which correspond to the two cases in Eq. \eqref{eq:k1z}. Eq. \eqref{eq:tm1} provides a robust lower limit on the duration of the sub-bursts that can be produced by self-modulation.

In the case of the repeating FRB 121102, the observed sub-burst duration is $\sim 0.5-1{\rm\; ms}$, and it is anti-correlated with the center frequency of the bursts \citep[][]{Hessels+2019}, which is consistent with Eq. \eqref{eq:tm2}. However, the observed sub-burst durations require the plasma density to be significantly larger than our fiducial value. Sub-bursts of finite duration have been observed also in other FRBs, including the FRB 121002 \citep[][]{Champion+2016}, the FRB 170827 \citep[][]{Farah+2018}, the FRB 181017 \citep[][]{Farah+2019}, the FRB 181112 \citep[][]{Cho+2020}, and the repeating FRB 180814.J0422+73 \citep[][]{Chime2019}. The shortest observed sub-burst duration, which is of the order of $10{\rm\; \mu s}$ for the FRBs 170827 and 181112, is consistent with being produced by self-modulation, and does not necessarily imply an upper limit on the duration of the burst.

Finally, note that our model does not explain the frequency drift observed in the repeating FRBs 121102 and 180814.J0422+73 (a similar drift has been also detected in other repeaters by \citealt{Chime2019a}). The frequency drift may be produced inside the source \citep[e.g.][]{Beloborodov2017, Beloborodov2020, Metzger+2019, Lyutikov2020}.

\section{Conclusions}
\label{sec:conclusions}

We have studied the possible effects of self-modulation on FRBs by considering the propagation of a monochromatic linearly-polarised wave with frequency $\nu_0\sim 1{\rm\; GHz}$ through a uniform plasma slab of density $N$, located at distance $R$ from the source. Strong self-modulation occurs if its growth rate $\Gamma$ exceeds $\sim 10\; c/R$ (then a seed perturbation is amplified by $\gtrsim 10$ e-foldings as the wave crosses the slab). The condition that $\Gamma R/c\gtrsim 10$ requires the plasma slab to be located within a critical radius $R_{\rm crit}\sim 10^{17}(N/10^2{\rm\; cm}^{-3})(L/10^{42}{\rm erg\; s}^{-1}){\rm\; cm}$, where $L$ is the isotropic equivalent of the FRB luminosity. Self-modulation breaks the burst into pancakes transverse to the radial direction. The observational signature that self-modulation leaves on FRBs depends on the position of the plasma slab:
\begin{itemize}
\item If $R\lesssim R_{\rm crit}$, the transverse size of the pancakes is smaller than the Fresnel scale. The pancakes are strongly broadened by diffraction as the burst exits the plasma slab, and the observer sees the interference pattern of a large number of pancakes. Interference produces a broadband frequency modulation of the burst, with bandwidth $\Delta\nu\sim 0.6(R/R_{\rm crit})\nu_0$. This effect is illustrated in Figure \ref{fig:1}.
\item If $R\sim R_{\rm crit}$, the transverse size of the pancakes is comparable with the Fresnel scale. Hence, the time structure produced by self-modulation is not smeared out due to diffraction. The observed intensity of the burst is modulated on a timescale of ten microseconds, which corresponds to the radial width of the pancakes. This effect is illustrated in Figure \ref{fig:2}.
\end{itemize}
Since in reality the plasma distribution along the line of sight is likely continuous, the natural next step is to consider the propagation of the FRB through a sequence of plasma slabs. We speculate that propagation at $R\lesssim R_{\rm crit}$ generates frequency modulation, and then a strong temporal structure (sub-bursts) develops at $R\sim R_{\rm crit}$, before self-modulation stops affecting the wave. This may explain the time-frequency structure reported in FRB 121102 \citep[][]{Hessels+2019}. However, our model does not explain the origin of the observed frequency drift.

Several aspects of self-modulation are left for future investigation, including the effects of (i) continuous plasma distribution along the line of sight, (ii) strong plasma magnetisation, (iii) different plasma composition (electron-positron instead of electron-ion), and (iv) relativistic electron temperature. These effects may be particularly important closer to the source. Yet more challenging is the full analysis of self-modulation at small radii where the wave has strength parameter $a_0\gg 1$.

\section*{Acknowledgements}

We thank the anonymous referee for constructive comments and suggestions that improved the paper. YL acknowledges support from the German-Israeli Foundation for Scientific Research and Development grant I-1362-303.7/2016, and the Israeli Science Foundation grant 2067/19. AMB acknowledges support from NASA grant NNX17AK37G, NSF grant AST 2009453, the Simons Foundation grant \#446228, and the Humboldt Foundation. LS acknowledges support from the Sloan Fellowship, the Cottrell Scholar Award, DoE DE-SC0016542,  NASA ATP 80NSSC18K1104, and NSF PHY-1903412.

\section*{Data availability}

No new data were generated or analysed in support of this research.

\def\aap{A\&A}\def\aj{AJ}\def\apj{ApJ}\def\apjl{ApJ}\def\mnras{MNRAS}\def\prl{Phys. Rev. Lett.}
\def\araa{ARA\&A}\def\physrep{PhR}\def\sovast{Sov. Astron.}\def\nar{NewAR}\def\pasa{PASA}
\def\aapr{Astronomy \& Astrophysics Review}\def\apjs{ApJS}\def\nat{Nature}\def\na{New Astron.}
\def\prd{Phys. Rev. D}\def\pre{Phys. Rev. E}\def\pasp{PASP}
\bibliographystyle{mn2e}
\bibliography{2d}

\appendix

\section{Scattering time}
\label{sec:scattering}

We discuss the observational signatures of the modes described by Eqs. \eqref{eq:k2y}-\eqref{eq:g2}. These modes have a typical wave number $k_y\simeq \min[a_0\omega_{{\rm P}i}/c_{\rm s},\; \sqrt{a_0\omega_0 \omega_{{\rm P}i}}/c]$ in the transverse direction and $k_z\simeq a_0\omega_{{\rm P}i}/c$ in the longitudinal direction, and their growth rate is $\Gamma\simeq \min[a_0^2(c^2/c_{\rm s}^2)(\omega_{{\rm P}i}^2/\omega_0),\; a_0\omega_{{\rm P}i}]$. Since $k_y\gg k_z$, the instability breaks the wave packet into filaments elongated in the direction of propagation of the pump wave.

The instability may develop if the radial width of the pancakes, $\lambda_z\simeq 2\pi/k_z$, is shorter than the length $c\tau$ of the burst, where $\tau\sim 1{\rm\; ms}$. The condition that $\lambda_z\lesssim c\tau$ may be presented as $R\lesssim R_{\rm crit}$, where
\begin{align}
\nonumber
R_{\rm crit}\sim 2\times 10^{-5} & \left(\frac{S_{\nu_0}}{{\rm Jy}}\right)^{1/2} \left(\frac{\nu_0}{{\rm GHz}}\right)^{-1/2} \times \\
& \times \left(\frac{D}{{\rm Gpc}}\right) \left(\frac{N}{10^2{\rm\; cm}^{-3}}\right)^{1/2} \left(\frac{\tau}{{\rm ms}}\right) {\rm\; pc} \;.
\end{align}
Hence, the instability may develop only close to the source (our analysis remains valid since $a_0\lesssim 1$ at $R\sim R_{\rm crit}$). The modes described by Eqs. \eqref{eq:k2y}-\eqref{eq:g2} may play a dominant role at these small radii, since they have a larger growth rate and a shorter transverse size than the modes described by Eqs. \eqref{eq:k1y}-\eqref{eq:g1}.

Considering the effect of a plasma slab at $R\sim R_{\rm crit}$, whose thickness is slightly smaller than $R$, we find that $\lambda_y\ll\sqrt{\lambda_0 R}$. The scattering angle is $\theta_{\rm scat}\simeq \lambda_0/\lambda_y$, and the corresponding scattering time is $\tau_{\rm scat}\simeq R\theta_{\rm scat}^2/c$. If the plasma is hot, we find that
\begin{align}
\nonumber
\tau_{\rm scat} \sim 0.5 & \left(\frac{S_{\nu_0}}{{\rm Jy}}\right)^{1/2} \left(\frac{\nu_0}{{\rm GHz}}\right)^{-5/2} \left(\frac{D}{{\rm Gpc}}\right) \times \\
\label{eq:tau1}
& \times \left(\frac{N}{10^2{\rm\; cm}^{-3}}\right)^{1/2} \left(\frac{\tau}{{\rm ms}}\right)^{-1} \left(\frac{T}{10^7{\rm\; K}}\right)^{-1} {\rm\; ms} \;.
\end{align}
If the plasma is cold, we find that
\begin{equation}
\label{eq:tau2}
\tau_{\rm scat} \sim 2 \left(\frac{S_{\nu_0}}{{\rm Jy}}\right)^{1/2} \left(\frac{\nu_0}{{\rm GHz}}\right)^{-3/2} \left(\frac{D}{{\rm Gpc}}\right) \left(\frac{N}{10^2{\rm\; cm}^{-3}}\right)^{1/2} {\rm\; ms} \;.
\end{equation}
In general, $\tau_{\rm scat}$ will be the minimum of the two. Eqs. \eqref{eq:tau1}-\eqref{eq:tau2} correspond to the two cases in Eq. \eqref{eq:k2y}.

Eqs. \eqref{eq:tau1}-\eqref{eq:tau2} may be used to constrain the properties of the circumburst medium by requiring that the contribution of self-modulation to the scattering time is shorter than a few milliseconds, which is the observed scattering time at the frequency of $1{\rm\; GHz}$ \citep[e.g.][]{CordesChatterjee2019}. However, two important caveats are (i) the fact that we have neglected the effect of the plasma magnetisation, which may be large in the region close to the source; (ii) the possible presence of pairs (electron-positron instead of electron-ion plasma).

\section{Derivation of the equations}
\label{sec:derivation}

\subsection{Derivation of Eqs. (7)-(8)}

We write the perturbed vector potential as $\delta a = \int \delta a(\omega',k_y',k_z')\exp[\i(\omega't-k_y'y-k_z'z)]{\rm d}\omega'{\rm d}k_y'{\rm d}k_z'$, and we introduce analogous definitions for the velocity perturbations and for the density perturbations. Substituting these definitions and Eqs. \eqref{eq:Ne}-\eqref{eq:ve} into Eq. \eqref{eq:deltaa}, and neglecting terms of order higher than $a_0^2$, we obtain
\begin{align}
\nonumber
\int\left(\omega'^2-c^2k'^2\right) & \delta a\left(\omega',k_y',k_z'\right) -\omega_{{\rm P}e}^2\int \delta v_{ex}\left(\omega',k_y',k_z'\right) = \\
\nonumber
= & -a_0 \omega_{{\rm P}e}^2\sin\chi_0 \int\delta n_e\left(\omega',k_y',k_z'\right) +\\
\label{eq:B1}
& -\frac{1}{4}a_0^2\omega_{{\rm P}e}^2\cos\left(2\chi_0\right) \int \delta v_{ex}\left(\omega',k_y',k_z'\right) \;,
\end{align}
where a factor of $\exp[\i(\omega't-k_y'y-k_z'z)]{\rm d}\omega'{\rm d}k_y'{\rm d}k_z'$ is implicit in all the integrals. Using the identities presented in Appendix \ref{sec:identities}, Eq. \eqref{eq:B1} gives
\begin{align}
\nonumber
& \left(\omega'^2-c^2k'^2\right) \delta a\left(\omega',k_y',k_z'\right) -\omega_{{\rm P}e}^2 \delta v_{ex}\left(\omega',k_y',k_z'\right) = \\
\nonumber
= & -\frac{\i}{2}a_0 \omega_{{\rm P}e}^2 \delta n_e\left(\omega'+\omega_0,k_y',k_z'+k_0\right) +\\
\nonumber
& + \frac{\i}{2}a_0 \omega_{{\rm P}e}^2 \delta n_e\left(\omega'-\omega_0,k_y',k_z'-k_0\right) +\\
\nonumber
& -\frac{1}{8}a_0^2\omega_{{\rm P}e}^2\delta v_{ex}\left(\omega'+2\omega_0,k_y',k_z'+2k_0\right) + \\
\label{eq:B2}
& -\frac{1}{8}a_0^2\omega_{{\rm P}e}^2 \delta v_{ex}\left(\omega'-2\omega_0,k_y',k_z'-2k_0\right) \;.
\end{align}
Substituting $\omega'=\omega+\omega_0$, $k_y'=k_y$, and $k_z'=k_z+k_0$ into Eq. \eqref{eq:B2}, we obtain Eq. \eqref{eq:deltaa+}. Substituting $\omega'=\omega-\omega_0$, $k_y'=k_y$, and $k_z'=k_z-k_0$, we obtain Eq. \eqref{eq:deltaa-}. We neglect terms proportional to $\delta v_{ex\pm 3}=\delta a_{\pm 3}$, which would give corrections of order higher than $a_0^4$ to the dispersion relation.

\subsection{Derivation of Eq. (12)}

Neglecting the relativistic corrections to the electron motion, the Euler's equation for the electron fluid is
\begin{align}
\nonumber
\frac{\pa {\bf V}_e}{\pa t} + \left({\bf V}_e\cdot\nabla\right) & {\bf V}_e = -\frac{m_i}{m_e} c_{\rm s}^2\nabla n_e+\\
\label{eq:B3}
& + \frac{e}{m_e}\left[\nabla\phi+\frac{1}{c}\frac{\pa{\bf A}}{\pa t}-\frac{{\bf V}_e}{c}\times\left(\nabla\times{\bf A}\right)\right] \;.
\end{align}
Taking into account that $\nabla n_e$ and $\nabla\phi$ are small, the zeroth order solution of Eq. \eqref{eq:B3} is ${\bf V}_e=e{\bf A}/m_ec$. Substituting such solution back into Eq. \eqref{eq:B3}, we find that
\begin{equation}
\label{eq:B4}
\frac{\pa {\bf V}_e}{\pa t} = -\frac{m_i}{m_e} c_{\rm s}^2\nabla n_e + \frac{e}{m_e}\nabla\phi+\frac{e}{m_ec}\frac{\pa{\bf A}}{\pa t} -\frac{e^2}{2m_e^2c^2}\nabla {\bf A}^2\;.
\end{equation}
Perturbing Eq. \eqref{eq:B4}, and neglecting terms that are quadratic in the perturbed quantities, we find that
\begin{equation}
\label{eq:B5}
\frac{\pa {\bf \delta V}_e}{\pa t} = -\frac{m_i}{m_e} c_{\rm s}^2\nabla\delta n_e + \frac{e}{m_e}\nabla\delta\phi+\frac{e}{m_ec}\frac{\pa\delta {\bf A}}{\pa t} -\frac{e^2}{m_e^2c^2}\nabla \left({\bf A}\cdot\delta {\bf A}\right) \;.
\end{equation}
Substituting ${\bf A}\cdot\delta {\bf A}=-(m_e^2c^4/e^2)a_0\sin\chi_0\delta a$ into Eq. \eqref{eq:B5}, we obtain Eq. \eqref{eq:eulere}.

\subsection{Derivation of Eqs. (13)-(15)}

Substituting Eq. \eqref{eq:conte} into the divergence of Eq. \eqref{eq:eulere}, and using the perturbed Gauss law, $\nabla^2 \delta \phi=4\pi N e(\delta n_e-\delta n_i)$, we find that
\begin{align}
\nonumber
\left(\frac{m_i}{m_e}c_{\rm s}^2\nabla^2\right. & -\left. \frac{\pa^2}{\pa t^2}\right) \delta n_e = \omega_{{\rm P}e}^2\left[\delta n_e-\delta n_i\right] + c^2a_0\sin\chi_0\nabla^2\delta a + \\
\label{eq:B6}
& -2c^2k_0a_0\cos\chi_0\frac{\pa}{\pa z}\delta a -c^2k_0^2a_0\sin\chi_0\delta a \;.
\end{align}
We write the perturbed vector potential as $\delta a = \int \delta a(\omega',k_y',k_z')\exp[\i(\omega't-k_y'y-k_z'z)]{\rm d}\omega'{\rm d}k_y'{\rm d}k_z'$, and we introduce an analogous definition for the density perturbations. Using the identities presented in Appendix \ref{sec:identities}, Eq. \eqref{eq:B6} gives
\begin{align}
\nonumber
& \left(\omega'^2 - \omega_{{\rm P}e}^2 - \frac{m_i}{m_e}c_{\rm s}^2k'^2 \right) \delta n_e\left(\omega',k_y',k_z'\right) + \omega_{{\rm P}e}^2 \delta n_i\left(\omega',k_y',k_z'\right) = \\
\nonumber
= & -\frac{\i}{2}a_0c^2 \left(k'+k_0\right)^2\delta a\left(\omega'+\omega_0,k_y',k_z'+k_0\right) + \\
\nonumber
& + \frac{\i}{2}a_0c^2 \left(k'-k_0\right)^2\delta a\left(\omega'-\omega_0,k_y',k_z'-k_0\right) + \\
\nonumber
& + \i a_0c^2 k_0 \left(k'_z+k_0\right)\delta a\left(\omega'+\omega_0,k_y',k_z'+k_0\right) + \\
\nonumber
& + \i a_0c^2 k_0 \left(k'_z-k_0\right)\delta a\left(\omega'-\omega_0,k_y',k_z'-k_0\right) + \\
\nonumber
& - \frac{\i}{2} a_0c^2 k_0^2\delta a\left(\omega'+\omega_0,k_y',k_z'+k_0\right) + \\
\label{eq:B7}
& + \frac{\i}{2} a_0c^2 k_0^2\delta a\left(\omega'-\omega_0,k_y',k_z'-k_0\right) \;.
\end{align}
Substituting $\omega'=\omega$, $k_y'=k_y$, and $k_z'=k_z$ into Eq. \eqref{eq:B7}, we obtain Eq. \eqref{eq:deltane0}. Substituting $\omega'=\omega+2\omega_0$, $k_y'=k_y$, and $k_z'=k_z+2k_0$, we obtain Eq. \eqref{eq:deltane+2}. Substituting $\omega'=\omega-2\omega_0$, $k_y'=k_y$, and $k_z'=k_z-2k_0$, we obtain Eq. \eqref{eq:deltane-2}. We neglect terms proportional to $\delta a_{\pm 3}$.

\section{Useful identities}
\label{sec:identities}

Suppose that $f= \int \tilde{f}(\omega',k_y',k_z')\exp[\i(\omega't-k_y'y-k_z'z)]{\rm d}\omega'{\rm d}k_y'{\rm d}k_z'$. The following identities turn out to be useful:
\begin{align}
\cos\left(m\chi_0\right)f = & \frac{1}{2}\int \exp\left[\i\left(\omega't-k_y'y-k_z'z\right)\right] \left[\tilde{f}_{+m}+\tilde{f}_{-m}\right]\\
\sin\left(m\chi_0\right)f = & \frac{\i}{2}\int\exp\left[\i\left(\omega't-k_y'y-k_z'z\right)\right] \left[\tilde{f}_{+m}-\tilde{f}_{-m}\right] \\
\nonumber
\cos\left(m\chi_0\right)\frac{\pa f}{\pa z} = & -\frac{\i}{2}\int\exp\left[\i\left(\omega't-k_y'y-k_z'z\right)\right] \times \\
& \times \left[\left(k_z'+mk_0\right)\tilde{f}_{+m}+\left(k_z'-mk_0\right)\tilde{f}_{-m} \right] \\
\nonumber
\sin\left(m\chi_0\right)\nabla^2f = & -\frac{\i}{2}\int\exp\left[\i\left(\omega't-k_y'y-k_z'z\right)\right] \times \\
& \times \left[\left(k'+mk_0\right)^2\tilde{f}_{+m}-\left(k'-mk_0\right)^2\tilde{f}_{-m}\right]
\end{align}
where we have defined $\tilde{f}_{\pm m}=\tilde{f}(\omega'\pm m\omega_0,k_y',k_z'\pm mk_0)$. All the integrals are performed over ${\rm d}\omega'{\rm d}k_y'{\rm d}k_z'$.

\end{document}